\documentclass[proceedings]{stacs}
\stacsheading{2009}{265--276}{Freiburg}
\firstpageno{265}

\theoremstyle{plain}
\theoremstyle{definition}

\usepackage{epsfig}
\usepackage{color}
\usepackage{amsmath}
\usepackage{amssymb}

\usepackage{verbatim}
\usepackage{graphicx,color, algorithmic, algorithm}
\begin{document}

\title[$l_{\infty}$-Fitting Robinson structures to distances]{an approximation
algorithm for $l_{\infty}$-fitting  Robinson structures to
distances}

\author[lab1]{V. Chepoi}{V. Chepoi}
\address[lab1]{LIF, Universit\'e d'Aix-Marseille, Marseille Cedex 9, France}
\email{chepoi,seston@lif.univ-mrs.fr}
\thanks{We are grateful to Bernard
Fichet for numerous insightful discussions during the work on this
paper.}
\thanks{The authors were partly supported by the ANR grant
BLAN06-1-138894 (projet OPTICOMB)}

\author[lab1]{M. Seston}{M. Seston}


\begin{abstract}

In this paper, we present a factor 16 approximation algorithm for
the following NP-hard distance fitting problem: given a finite set
$X$ and a distance $d$ on $X$,  find a Robinsonian distance $d_R$ on
$X$ minimizing the $l_{\infty}$-error
$||d-d_R||_{\infty}=\mbox{max}_{x,y\in X}\{ |d(x,y)-d_R(x,y)|\}.$ A
distance  $d_R$ on a finite set $X$ is Robinsonian if its matrix can
be symmetrically permuted so that its elements do not decrease when
moving away from the main diagonal along any row or column.
Robinsonian distances generalize ultrametrics, line distances and
occur in the seriation problems and in classification.

\end{abstract}

\maketitle

\vspace{-0.4cm}
\section{Introduction}\label{S:one}

\noindent{\sc\bf 1.1. Seriation problem.} Many applied algorithmic
problems involve ordering of a set of objects so that closely
coupled objects are placed near each other. These problems occur in
such diverse applications as data analysis,  archeological dating,
numerical ecology, matrix visualization methods, DNA sequencing,
overlapping clustering, graph linear arrangement, and sparse matrix
envelope reduction. For example, a major issue in classification and
data analysis is to visualize simple geometrical and relational
structures between objects. Necessary for such an analysis is a
dissimilarity on a set of objects, which is measured directly or
computed from a data matrix. The classical {\it seriation problem}
\cite{Hu,Ke2} consists in finding of a simultaneous permutation of
the rows and the columns of the dissimilarity matrix with the
objective of revealing an underlying one-dimensional structure. The
basic idea is that small values should be concentrated around the
main diagonal as closely as possible, whereas large values should
fall as far from it as possible. This goal is best achieved by
considering the so-called {\it Robinson property} \cite{Ro}: a
dissimilarity matrix has this property if its values do not decrease
when moving away from the main diagonal along any row or column.
Experimental data usually contain errors, whence the dissimilarity
can be measured only approximatively. As a consequence, any
simultaneous permutation of the rows and the columns of the
dissimilarity matrix gives a matrix which fails to satisfy the
Robinson property, and we are led to the problem of finding a
matrix reordering which is as close as possible to a Robinson
matrix. As an error measure one can use the $l_p$-distance
between two matrices.  Several heuristics for seriation using
Robinson matrices have been
considered in the literature  (the
package {\sf seriation} \cite{HaHoBu} contains their implementation).
However, these methods either have
exponential complexity or do not provide any optimality guarantee of
the obtained solutions. In this paper, we provide a factor 16
algorithm for the NP-hard problem of optimally fitting a dissimilarity
matrix by a Robinson matrix under the $l_{\infty}$-error.

\medskip\noindent{\sc\bf 1.2. Definitions and the problem.} Let $X$
be a set of $n$ elements to sequence, endowed with a {\it
dissimilarity function} $d:X^2\rightarrow {\mathbb R}^+\cup\{ 0\}$
(i.e., $d(x,y)=d(y,x)\ge 0$ and $d(x,x)=0$).  A dissimilarity $d$
and a total order $\prec$ on  $X$ are  {\it compatible} if
$d(x,y)\ge d(u,v)$ for any four elements such
that $x\prec u\prec v\prec y.$ Then $d$ is {\it Robinsonian} if it
admits a compatible order.
Basic examples of Robinson dissimilarities are the {\it
ultrametrics} and the standard {\it line-distance} between $n$
points on the line. Denote by ${\mathcal D}$ and ${\mathcal R}$ the
sets of all dissimilarities and of
all Robinson dissimilarities on $X.$   For
$d,d'\in {\mathcal D},$  define the $l_{\infty}$-error   by
$||d-d'||_{\infty}=\mbox{max}_{x,y\in X}\{ |d(x,y)-d'(x,y)|\}.$ To
formulate the corresponding fitting problem, we relax the notions of
compatible order and Robinson dissimilarity. Given $\epsilon\ge 0,$
a total order $\prec$ on $X$ is called $\epsilon$-{\it compatible}
if $x\prec u\prec v\prec y$ implies $d(x,y)+2\epsilon\ge d(u,v).$ 
An $\epsilon$-{\it Robinsonian dissimilarity} is a
dissimilarity admitting an $\epsilon$-compatible order, i.e., for
each pair $x,y\in X$ one can pick a value $d_R(x,y)\in
[d(x,y)-\epsilon, d(x,y)+\epsilon]$ so that the resulting
dissimilarity $d_R$ is Robinsonian. In this paper, we study the
following NP-hard \cite{ChFiSe} optimization problem:

\medskip\noindent {\small\bf Problem
$l_{\infty}$-FITTING-BY-ROBINSON:} {\it Given $d\in
{\mathcal D},$ find a Robinson dissimilarity $d_R\in {\mathcal R}$
minimizing the $l_{\infty}$-error $||d-d_R||_{\infty},$  i.e., find
a least $\epsilon$ such that $d$ is $\epsilon$-Robinsonian.}

\medskip\noindent {\sc\bf 1.3. Related work.} Fitting general
distances by simpler distances (alias low-distortion embeddings) is
a classical problem in mathematics, data analysis, phylogeny, and,
more recently, in computer science.  We review here only the results
about $l_{\infty}$-fitting of distances (this error measure is also
known as the {\it maximum additive distortion} or
the {\it maximum additive two-sided error} \cite{BaDhGuRaRaRaSi}).
Farach et al. \cite{FaKaWa} showed that $l_{\infty}$-fitting of a
distance $d$ by an ultrametric is polynomial. This result has been
used by Agarwala et al. \cite{AgBaFaNaPa} to design a factor 3
approximation algorithm for $l_{\infty}$-fitting of  distances by
tree-distances, a problem which has been shown to be strongly
NP-hard \cite{AgBaFaNaPa}. A unified and simplified treatment of
these results of \cite{AgBaFaNaPa,FaKaWa} using sub-dominants was
given in \cite{ChFi2}.  A factor 2 approximation algorithm for the
NP-hard problem of $l_{\infty}$-fitting of a dissimilarity by a
line-distance was given by H\symbol{229}stad et al. \cite{HaIvLa}.
B\u{a}doiu \cite{Ba} proposed a constant-factor algorithm for
$l_{\infty}$-fitting of distances by $l_1$-distances in the plane.

Seriation is important in archeological dating,
clustering hypertext orderings, numerical ecology, sparse matrix ordering, matrix
visualization methods, and DNA sequencing
\cite{AtBoHe,CaPi,Hu,Ke2,MiRo,Ro}. A package {\sf seriation} implementing  various
seriation methods is described  in \cite{HaHoBu}. The most common
methods for clustering provide a
visual display of data in the form of dendrograms. Dissimilarities
in perfect agreement with dendrograms (i.e.,
ultrametrics) are Robinsonian. Generalizing this correspondence, \cite{Di,DuFi}
establish that the Robinson dissimilarities
can be visualized by hierarchical structures called pyramids.

\medskip\noindent {\sc\bf 1.4. Our result and techniques.} The main
result of the paper is a factor 16 approximation algorithm for the
problem $l_{\infty}$-{\sc FITTING-BY-ROBINSON}. The basic setting of
our algorithm goes as follows. First we show that the optimal  error
$\epsilon^*$  belongs to a
well-defined list $\Delta$ of size $O(n^4).$ As in some other minmax
problems, our approximation algorithm tests the entries of
$\Delta$, using a parameter $\epsilon$, which is the ``guess" for $\epsilon^*.$
For current  $\epsilon\in \Delta,$ the algorithm
either finds that no $\epsilon$-compatible order exist, in which
case the input dissimilarity $d$ is not $\epsilon$-Robinsonian, or
it returns a $16\epsilon$-compatible order. Now, if $\epsilon$ is
the least value for which the algorithm does not return the negative
answer, then $\epsilon^*\ge \epsilon$, and the returned
$16\epsilon$-Robinsonian dissimilarity  has $l_{\infty}$-error at
most $16\epsilon^*,$ establishing that we have a factor $16$
approximation algorithm.

For $\epsilon\in \Delta,$ a canonical binary relation $\preccurlyeq$
is computed so that any $\epsilon$-compatible total order refines
$\preccurlyeq$ or its dual. If $\preccurlyeq$ is not a partial
order, then the algorithm halts and returns the negative answer. If
$\preccurlyeq$ is a total order, then  we are done. Otherwise, we
select a maximal chain $P=(a_1,a_2,\ldots,a_p)$ of the partial order
$\preccurlyeq$ and search to fit each element of
$X^{\circ}:=X\setminus P$ between two consecutive elements of  $P.$
We say that  $a_i,a_{i+1}\in P$ form a {\it hole} $H_i$ and that all
elements $x\in X^{\circ}$ assigned between $a_i$ and $a_{i+1}$ are
{\it located} in $H_i.$   This distribution of the elements to holes
is performed so that (a) all elements $X_i$ of $X^{\circ}$ located
in the same hole $H_i$ must ``fit" in this hole, i.e., for all
$x,y\in X_i$ one of the orders $a_i\prec x\prec y\prec a_{i+1}$ or
$a_i\prec y\prec x\prec a_{i+1}$ must be $c\epsilon$-compatible for
some $c\le 12.$ Partitioning $X^{\circ}$ into sets $X_i,$
$i=1,\ldots,p-1,$ is not obvious. Even if such a partition is
available, we cannot directly apply a recursive call to each $X_i$,
because (b) the elements located outside the hole $H_i$ will impose
a certain order on the elements of $X_i$ and, since we tolerate some
errors, (c) we cannot ensure that $X_i$ is exactly the set of
elements which must be located in $H_i$ in some
$\epsilon$-compatible total order. To deal with (a), we give a
classification of admissible and pairwise admissible holes for
elements of $X^{\circ}.$ This allows to show that, if we tolerate a
$12\epsilon$-error, then each element $x\in X^{\circ}$ can be
located in the leftmost or rightmost admissible hole for $x$ (we
call them {\it bounding holes} of $x$). Both locations are feasible
unless several elements have the same pair of bounding holes. For
$i<j,$ let $X_{ij}$ be the set of all elements of $X^{\circ}$ having
$H_i$ and $H_{j-1}$ as bounding holes. To deal with (b) and (c), on
each set $X_{ij}$ we define a directed graph ${\mathcal
L}_{ij}^{\rightarrow}.$  The strongly connected components (which we
call {\it cells}) of ${\mathcal L}_{ij}^{\rightarrow}$ have the
property that in any $\epsilon$-compatible order all elements of the
same component must be located in the same hole. In fact the cells
(and not the sets $X_i$) are the units to which we apply the
recursive calls in the algorithm. To decide in which hole $H_i$ or
$H_{j-1}$ to locate each cell of ${\mathcal L}_{ij}^{\rightarrow}$
and to define the relative order between the cells assigned to the
same hole, we define another directed graph ${\mathcal G}_{ij}$
whose vertices are the cells of ${\mathcal L}_{ij}^{\rightarrow}$ in
such a way that (i) if some ${\mathcal G}_{ij}$ does not admit  a
partition into two acyclic subgraphs then no $\epsilon$-compatible
order exist and (ii) if ${\mathcal G}_{ij}$ has a partition into two
acyclic subgraphs ${\mathcal G}^-_{ij}$ and ${\mathcal G}^+_{ij},$
then all cells of ${\mathcal G}^-_{ij}$ will be located in $H_i,$
all cells of ${\mathcal G}^+_{ij}$ will be located in $H_{j-1},$ and
the topological ordering of each of these graphs defines the
relative order between the cells. To partition ${\mathcal G}_{ij}$
into two acyclic subgraphs (this problem in general is NP-complete
\cite{John}), we investigate the specific properties of graphs in
question, allowing us to define a 2-SAT formula $\Phi_{ij}$ which is
satisfiable  if and only if the required bipartition of ${\mathcal
G}_{ij}$ exists. Finally, to locate in each hole $H_i$ the cells
coming from different subgraphs ${\mathcal G}^+_{j'i},{\mathcal
G}^-_{ij},$ and ${\mathcal G}^-_{ij''}$ with $j'<i<j<j'',$ we use
the following separation rule: the cells of ${\mathcal G}^+_{j'i}$
are located to the left of the cells of ${\mathcal G}^-_{ij}$ and
the cells of ${\mathcal G}^-_{ij}$ are located to the right of the
cells of ${\mathcal G}^-_{ij''}.$ Due to space
constraints, all missing proofs are given in the full
version \cite{ChSe}.

\vspace{-0.4cm}
\section{Preliminary results}

The $\prec$-restricted problem is
obtained from {\sc $l_{\infty}$-FITTING-BY-ROBINSON} by fixing the total order $\prec$ on $X.$
Let $\check{d}_{\prec}$ be
a dissimilarity defined by setting
$\check{d}_{\prec}(x,y)=\max\{d(u,v): x \prec u \prec v \prec y\}$
for all $x,y\in X$ with  $x \prec y$ (we suppose here that $a\prec
a$ for any $a\in X$). Let
$2\tilde{\epsilon}_{\prec}=||d-\check{d}_{\prec}||_{\infty}$ and let
$\tilde{d}_{\prec}$ be the (Robinsonian) dissimilarity obtained from
$\check{d}_{\prec}$ by setting $\tilde{d}_{\prec}(x,y)=\max
\{\check{d}_{\prec}(x,y)-\tilde{\epsilon}_{\prec}, 0\}$ for all
$x,y\in X, x\ne y.$ Then, the following holds:

\begin{proposition} \label{superdom} For a total
order $\prec$ on $X$ and $d\in {\mathcal D}$, $\tilde{d}_{\prec}$
minimizes  $|| d-d'||_\infty.$
\end{proposition}

Proposition \ref{superdom}  establishes that an optimal solution
of the problem {\sc $l_{\infty}$-FITTING-BY-ROBINSON} can be
selected among $n!$ Robinsonian dissimilarities of the form
$\tilde{d}_{\prec}.$ In the full version, we show that the natural
heuristic similar to the factor 3 approximation algorithms of
H{\aa}stad et al. \cite{HaIvLa} and Agarwala et al.
\cite{AgBaFaNaPa} (which instead of $n!$ total orders considers
only $n$ orders) does not provide a constant-factor approximation
algorithm for our problem. Proposition \ref{superdom} also implies
that the optimal error $\epsilon^*$ in {\sc
$l_{\infty}$-FITTING-BY-ROBINSON} belongs to a well-defined list
$\Delta=\{\frac{1}{2}|d(x,y)-d(x',y')|: x,y,x',y'\in X\}$ of size
$O(n^4).$

Given $d\in {\mathcal D}$ and  $\epsilon\in \Delta,$ we define a
partial order $\preccurlyeq$ such that every $\epsilon$-compatible
total order $\prec$ refines either $\preccurlyeq$ or its dual.
For this, we set $p\preccurlyeq q$ for two arbitrary elements
$p,q\in X,$ and close $\preccurlyeq$ using the properties of partial
orders and the following observation: {\it if $d(x,y)>\max\{
d(x,z),d(z,y)\}+2\epsilon,$ then in all $\epsilon$-compatible with
$d$ orders $z$ must be located between $x$ and $y.$} In this case,
if we know that two of the elements $x,z,y$ are in relation
$\preccurlyeq$ then we can extend this relation to the whole
triplet. For example, if we know that $x\preccurlyeq z,$ then we
conclude that also $z\preccurlyeq y$ and $x\preccurlyeq z$.
If the resulting $\preccurlyeq$ is
not a partial order, then $d$ does not admit an
$\epsilon$-compatible total order. So, further let $\preccurlyeq$ be
a partial order. For two disjoint subsets $A,B$ of $X,$ set
$A\preccurlyeq B$ if $a\preccurlyeq b$ for any $a\in A$ and $b\in
B.$  We write $x?y$ if neither $x\preccurlyeq y$ nor $y\preccurlyeq
x$ hold.  For two numbers $\alpha$ and $\beta$ we will use the
following notations (i) $\alpha\thickapprox_c\beta$ if
$|\alpha-\beta|\le c\epsilon$, (ii) $\beta\gtrsim_c \alpha$ if
$\beta\ge \alpha-c\epsilon$, and (iii) $\beta\gg_c \alpha$ if
$\beta>\alpha+c\epsilon$. We continue with basic properties of the
canonical partial order $\preccurlyeq$: {\it If  $w \preccurlyeq \{
v,z\},$ $v?z,$ $u \preccurlyeq v,$ $u?z,$ and $w?u,$
then: \textup{(i)} $d(v,w)\thickapprox_2d(z,w);$ \textup{(ii)} $d(v,z)\lesssim_2 \min\{ d(v,w),d(z,w)\}$;
\textup{(iii)} $d(w,z)\thickapprox_4 \{d(u,v),d(u,z)\};$ \textup{(iv)} $d(w,u) \lesssim_2 \min
\{d(w,v),d(u,v)\}$. }

\vspace{-0.4cm}
\section{Pairwise admissible holes}

\noindent{\sc\bf 3.1. Admissible holes.} Let
$P=(a_1,a_2,\ldots,a_{p-1},a_p)$ be a maximal chain of the partial
order  $\preccurlyeq$. For notational convenience, we assume that
all elements of $X^{\circ}$ must be located between $a_1$ and $a_p$
($a_1$ and $a_p$ can be artificially added); this way,
every element of $X^{\circ}$ must be located in a hole. Let $H_{ij}$
be the union of all holes comprised between $a_i,a_j.$ For  $x\in
X^{\circ},$ denote by $H(x)$ the union of all holes $H_i$ such that
$x?a_i$ or $x?a_{i+1}.$ If $H(x)=H_{ij},$ the holes $H_i$ and
$H_{j-1}$ are called {\it bounding holes}; see Fig. 1 (note that
$a_i=\max \{a_k \in P: a_k \preccurlyeq x\}$ and $a_j=\min \{a_k \in
P: x \preccurlyeq a_k\}$ for $x\in X^{\circ}).$ All other holes of
$H(x)$ are called {\it inner holes}.  Since $x\notin P,$ $H(x)$
contains at least two holes. The hole $H_k$ of $H(x)$ is $x$-{\it
admissible,} if the total order on $P\cup \{ x\}$ obtained from
$\preccurlyeq$ by adding the relation $a_k\preccurlyeq x\preccurlyeq
a_{k+1}$ is $\epsilon$-compatible with $d$. It can be easily shown
that the bounding holes of $H(x)$ must be $x$-admissible. Denote by
$d_x$ the mean value of $\min \{ d(x,a_k): i < k< j\}$  and $\max \{
d(x,a_k): i<k<j\}.$ We call
$\delta_k=d(a_k,a_{k+1})$ the {\it size} of the hole $H_k.$ Then the
following holds:

\begin{lemma} \label{sizehole} If  an inner hole $H_k$ of $H(x)$
is $x$-admissible, then $d_x\thickapprox_1
\{d(x,a_k),d(x,a_{k+1})\} $ $\thickapprox_2 \delta_k .$ In
particular, $\delta_k\thickapprox_3 d_x.$ More generally, for all
$k,k' \in ]i,j[,$ we have $d_x\gtrsim_3 d(a_k,a_{k'}).$
\end{lemma}

\medskip\noindent{\sc\bf 3.2. Pairwise admissible holes.}  A pair $\{
H_k,H_{k'}\}$ of holes is called $(x,y,c)$-{\it admissible} if
$H_k$ is $x$-admissible, $H_{k'}$ is $y$-admissible, and the total
order on $P\cup \{ x,y\}$ obtained  by adding to $\preccurlyeq$
the relations $a_k\preccurlyeq x\preccurlyeq a_{k+1}$ and
$a_{k'}\preccurlyeq y\preccurlyeq a_{k'+1}$ is
$c\epsilon$-compatible. Denote by $AH(x)$ the set of all
$x$-admissible holes $H_k$ so that for each  $y\in X^{\circ},$
$y\ne x$,  there exists an $y$-admissible hole $H_{k'}$ such that
$\{ H_k,H_{k'}\}$ is a $(x,y,1)$-admissible pair. Further we can
assume that for any $x\in X^{\circ}$ the bounding holes of
$H(x)=H_{ij}$ belong to $AH(x).$ Otherwise, if say $H_i\notin
AH(x),$ then $a_{i+1} \prec x$ in any $\epsilon$-compatible total
order $\prec$ extending $\preccurlyeq,$ thus we can augment the
canonical partial order $\preccurlyeq$ by setting
$a_{i+1}\preccurlyeq x$ and by reducing the segments $H(x)$
accordingly. Next we investigate the pairwise admissible locations
of $x$ and $y$ in function of the mutual geometric location of the
segments $H(x)$ and $H(y)$ and of the values $d(x,y),d_x,$ and
$d_y.$ We distinguish the following cases: {\bf (H1)} $H(x)=H(y);$ {\bf (H2)} $H(x)$ and
$H(y)$ are disjoint; {\bf (H3)} $H(x)$ and $H(y)$ overlap in at least 2 holes
($H(x)\circ H(y)$); {\bf (H4)} $H(x)$  and $H(y)$ overlap in a single hole
($H(x)\ast H(y)$); {\bf (H5)} $H(y)$ is a proper subinterval of $H(x)$
($H(y)\Subset H(x)$). This classification of pairs $\{ x,y\}$ of $X^{\circ}$ is
used in the design of our approximation algorithm. Also the proofs
of several results employ a case analysis based on (H1)-(H5). We
continue with the following result. It specifies the constraints
on pairs of elements,  each
element of $X^{\circ}$ can be located in one of its bounding holes.

\begin{proposition} \label{pairwise_admissible} For two elements $x,y\in X^{\circ}$, any
location of $x$ in a bounding hole of $H(x)=H_{ij}$ and any
location of $y$ in a bounding hole of $H(y)=H_{i'j'}$ is
$(x,y,12)$-admissible, unless $H(x)=H(y)$ and $d(x,y)\ll_3 \max\{
d_x,d_y\}$ or $d(x,y)\gg_3
\max\{d_x,d_y\},$ subject to the following three constraints: \textup{(i)} if $H(x) \Subset H(y),$ $x$ and $y$ are
located in a common bounding hole, then $x$ is  between $y$ and $a_{i+1};$
\textup{(ii)} if $H(x) \ast H(y),$ then $i <i'$ implies $x \prec y;$
\noindent\textup{(iii)} if $H(x)=H(y)$,  $x$ and $y$ are located
in the same bounding hole, and $d_y\ll_4 d_x,$ then $y$ is
between $x$ and  $a_{i+1}$. If $H(x)=H(y)$ and $d(x,y)\gg_3 \max\{ d_x,d_y\},$ then
the only $(x,y,1)$-admissible locations are the two locations of
$x$ and $y$ in different bounding holes. If $H(x)=H(y)$ and
$d(x,y)\ll_3 \max\{ d_x,d_y\},$ then any $(x,y,1)$-admissible
location is in common $x$- and $y$-admissible holes.
\end{proposition}

\vspace{-0.4cm}
\section{Distributing elements to holes}

In this section, we describe how,  for each hole $H_i$, to compute
the set $X_i$ of elements of $X^{\circ}$ which will be located in
$H_i.$ This set consists of some $x$ such that $H_i$ is a
bounding hole of $H(x).$ Additionally, each $X_i$ will be
partitioned into an ordered list of cells, to which we perform
recursive calls. Let $X_{ij}$ consist of all $x\in X^{\circ}$ such
that $H(x)=H_{ij}.$ The sets $X_{ij}$ form a partition of
$X^{\circ}.$ In the next subsections, we will show how to
partition each  $X_{ij}$ into two subsets $X_{ij}^-$ and $X_{ij}^+,$
so that $X_{ij}^-$ will be located in $H_i$ and $X_{ij}^+$ in
$H_{j-1};$ see Fig. 1.

\vspace*{-0.1cm}
\begin{figure}
\begin{center}
\scalebox{0.8}{\input{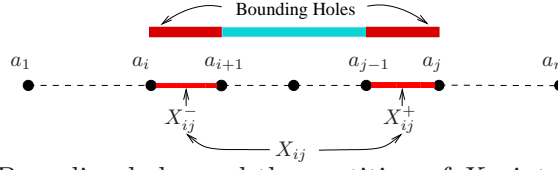}}
\end{center}
\vspace*{-0.55cm} \caption{Bounding holes and the partition of
$X_{ij}$ into $X_{ij}^-$ and $X_{ij}^-$} \vspace*{-0.35cm}
\end{figure}

\medskip\noindent{\sc\bf  4.1. Blocks, cells, and clusters.} Two
elements $x,y\in X_{ij}$ are called {\it linked} ({\it separated}) if in all
$(x,y,1)$-admissible locations $x$ and $y$ must be placed in the
same hole (in distinct  bounding holes). Two subsets $A$
and $B$ of $X_{ij}$ must be {\it separated} if all $x\in A$ and $y\in B$ are separated.
Let $S_{ij}$ and $L_{ij}$ be the sets of all pairs $x,y\in X_{ij}$ such that
 $d(x,y)\gg_3\mbox{max}\{ d_x,d_y\},$ resp., $d(x,y)\ll_3 \max\{
d_x,d_y\}.$ By Proposition \ref{pairwise_admissible}, all pairs of
$S_{ij}$ are separated  and  all pairs of $L_{ij}$ are linked.
Since ``be linked" is an equivalence relation, all vertices of the
same connected component (called {\it block}) of the graph ${\mathcal
L}_{ij}=(X_{ij},L_{ij})$ are linked.  We continue by
investigating in which cases two blocks of ${\mathcal L}_{ij}$ are
separated or linked. For  $x,y\in X_{ij},$ set $x\rightarrowtail
y$ iff {\bf (A1)} $d_x \ll_4 d_y$ or {\bf (A2)} $d_x \gtrsim_4
d_y$ and there exists  $z\in X_{ij}$ such that $xz,yz\notin
L_{ij}$ and $d(x,z) \ll_{16} d(y,z).$ If $x,y,z\in
X_{ij}$ satisfy {\bf (A2)}, then it can be shown that  $y$
and $z$ are {\it strongly
separated}, i.e., $d(y,z)\gg_9\mbox{max}\{ d_y,d_z\}.$ Additionally, we show that
if $x\rightarrowtail y$, then $x\prec
y$ in all  $\epsilon$-compatible orders $\prec$ such that $a_{i+1}
\prec \{ x,y\}$ and $y\prec x$ in all $\epsilon$-compatible orders
$\prec$ such that $\{ x,y\}\prec a_{j-1}$.

On $X_{ij}$ we define a directed graph
${\mathcal L}_{ij}^{\rightarrow}:$ we draw an arc $x\rightarrow y$
iff {\bf (L1)} $x\rightarrowtail y$ and $x,y$ belong to a common
block of ${\mathcal L}_{ij}$ or {\bf (L2)} $d(x,y)\ll_5 \max\{
d_x,d_y\}.$ If {\bf (L2)} is satisfied, then $xy\in L_{ij}$ and
$y\rightarrow x$ hold. The strongly connected components of
${\mathcal L}_{ij}^{\rightarrow}$ are called {\it cells}. Every block
is a disjoint union of cells. Indeed, if $x,y$
belong to a common cell, let $R$ be a directed path of ${\mathcal
L}_{ij}^{\rightarrow}$ from $x$ to $y.$ Pick any arc $u\rightarrow
v$ of $R$. If it has type {\bf (L2)}, then $uv\in {\mathcal
L}_{ij}.$ Otherwise, if $u\rightarrow v$ has type {\bf (L1)}, then
$u$ and $v$ belong to a common block. Thus the ends
of all arcs of any path between $x,y$ belong to a common
block.

\begin{lemma} \label{separate-between} Let $x,x',y\in X_{ij}.$ If
$x,x'$ belong to a common cell, but $\{ x,x'\}$ and $y$ belong to
distinct blocks, then there does not exist an $\epsilon$-compatible
order such that $x \prec y \prec x'.$
\end{lemma}

\begin{lemma} \label{R4} For cells $C',C'',$ if
 $x,x'\in C',$ $y,y'\in C'',$  and $x\rightarrowtail y,$
$y'\rightarrowtail x',$ then   $C'$ and  $C''$ must be separated.
\end{lemma}

\begin{proof}    Let $B',B''$ be the blocks containing $C',C''.$
If $B'=B'',$ as $x\rightarrowtail y$ and
$y'\rightarrowtail x',$ they are (L1)-arcs, hence $x\rightarrow y$
and $y'\rightarrow x'.$ This is impossible  since $\{ x,x'\}$
and $\{ y,y'\}$ belong to distinct cells. Thus $B'\ne B''.$ By
Lemma \ref{separate-between}, if we locate $x,x',y,y'$ in the same
bounding hole $H_j$, either $\{x,x'\} \prec \{y,y'\}$ or $\{y,y'\}
\prec \{x,x'\}$ holds. On the other hand,
$x\rightarrowtail y,$ $y'\rightarrowtail x'$ imply that $x \prec y$
and $y'\prec x'.$ Thus $C'$ and
$C''$ must be separated. \end{proof}

Now, let ${\mathcal S}_{ij}$ be a graph having cells as vertices and
an edge between two cells $C',C''$ iff {\bf (S1)}
there exist $x,y\in X_{ij},$ $x$ in the same block as $C'$
and $y$ in the same block as $C''$ such that $xy\in S_{ij}$ or {\bf
(S2)} there exist $x,x'$ in the same block as $C'$ and
$y,y'$ in the same block as $C''$ such that each pair
$xx'$ and $yy'$ belong to a common cell, and $x\rightarrowtail y,
y'\rightarrowtail x'.$ By Proposition \ref{pairwise_admissible} and
Lemma \ref{R4}, in cases {\bf (S1)} and {\bf (S2)} the sets
$C'$ and $C''$ must be separated. The graph ${\mathcal S}_{ij}$ must
be bipartite, otherwise no $\epsilon$-compatible order exist. Now,
for each connected component of ${\mathcal S}_{ij}$ consider its
canonical bipartition $\{ A',A''\}$, and draw an edge between any
two cells, one from $A'$ and another from $A''.$ Denote the obtained
graph also by ${\mathcal S}_{ij}.$ Call the union of  cells from
$A'$ (or from $A''$) a {\it cluster}. The clusters ${\mathcal K}'$
and ${\mathcal K}''$ of $A'$ and $A''$ are called {\it twins}. From
the construction, we immediately obtain that all elements of a
cluster are linked and two twin clusters are separated. A connected
bipartite component $\{ {\mathcal K}',{\mathcal K}''\}$ of
${\mathcal S}_{ij}$ is called a {\it principal component} if there
exists $x\in {\mathcal K}'$ and $y\in {\mathcal K}''$ such that $x$
and $y$ are strongly separated.

\medskip\noindent{\sc\bf  4.2. Partitioning  $X_{ij}$ into $X^-_{ij}$
and $X^+_{ij}.$} We describe how to partition  $X_{ij}$ into the
subsets $X^-_{ij}$ and $X^+_{ij}.$ For this, we define a directed
graph ${\mathcal G}_{ij}$ having cells as
vertices, and an arc $C' \rightarrowtail C$ with tail
$C'$ and head $C$ exists iff one of the following
conditions is satisfied: {\bf (G1)} $C'$ and $C$ belong to  twin clusters of
${\mathcal S}_{ij};$ {\bf (G2)} $C'$ and $C$ are not connected by (G1)-arcs and
there exist $x\in C$ and $x'\in C'$ such that $d_{x'} \ll_4 d_{x};$
{\bf (G3)} $C'$ and $C$ are not connected by (G1)- or
(G2)-arcs and there exist  $x\in C, x'\in C',$ and $z\in X_{ij}$
such that $xz,x'z\notin L_{ij}$ and $d(x',z) \ll_{16} d(x,z).$
A head of a (G3)-arc is called a (G3)-{\it cell}.  A (G$i$)-{\it
cycle} is a directed cycle of ${\mathcal G}_{ij}$ with arcs of type
{\bf (G$i$)}, $i=1,2,3.$ The (G1)-cycles are
exactly the cycles of length 2. A {\it mixed cycle} is a directed
cycle containing arcs of types {\bf (G2)} and {\bf (G3)}. Finally,
an {\it induced cycle} is a directed cycle $\mathcal C$ such that
for two cells $C,C'\in {\mathcal C}$ we have $C' \rightarrowtail C$
if and only if $C$ is the successor of $C'$ in ${\mathcal C}.$ Our
next goal is to establish that either the set of cells can be
partitioned into two subsets such that the subgraphs of ${\mathcal
G}_{ij}$ induced by these subsets do not contain directed cycles or
no $\epsilon$-compatible order exist. Deciding if a directed graph
can be partitioned into two acyclic subgraphs is NP-complete \cite{John}.
In our case,  this can be done in polynomial time by exploiting the
structure of ${\mathcal
G}_{ij}$.

\begin{lemma}\label{cycle} If ${\mathcal C}=(C_1,C_2,\ldots,C_k,C_1)$ is a directed  cycle of
${\mathcal G}_{ij}$, then for any $\epsilon$-compatible order,
$\mathcal C$ has a cell located in the hole $H_i$ and a cell located
in the hole $H_{j-1}.$
\end{lemma}

\begin{proof} The assertion
is obvious if $\mathcal C$  is a (G1)-cycle. So, suppose that all
arcs of $\mathcal C$ have type {\bf (G2)} or {\bf (G3)}. The
definition of cells implies that $\mathcal C$ contains two
consecutive cells, say $C_1$ and $C_k$, which belong to different
blocks. Suppose that there exists an $\epsilon$-compatible order
$\prec$ such that no element of $\cup_{l=1}^k C_l$ is located in the
hole $H_{i}=[a_{i},a_{i+1}]$, i.e., $a_{i+1}\prec \cup_{l=1}^k C_l$.
In each $C_l$ pick two elements $x_l,y_l$ such that
$x_l\rightarrowtail y_{l+1(mod k)}.$ Then  $x_l\prec y_{l+1(mod k)}$
for all $l=1,\ldots,k.$ We divide
the cells of $\mathcal C$ into groups: a group consists of
all consecutive cells of $\mathcal C$ belonging to one and the same
block. The first group starts with $C_1,$ while the last group ends
with $C_k.$ We assert that if $\{ C_{l-q},\ldots,C_l\}$ and $\{
C_{l+1},\ldots ,C_{l+r}\}$ are two consecutive groups of $\mathcal
C$, then $C_l\prec C_{l+1}\cup \cdots \cup C_{l+r}$ (all indices
here are modulo $k$). Indeed, pick $u\in C_l$ and $v\in C_{l+1}.$
Since $\{x_l,u\}$ and $\{y_{l+1},v\}$ belong to different blocks
while each of these pairs belong to a common cell, applying Lemma
\ref{separate-between} to each of the triplets of the quadruplet
$x_l,u,y_{l+1},v,$ we infer that in the total order $\prec$ none of
$y_{l+1},v$ is located between $x_l$ and $u$ and none of $x_l,u$ is
located between $y_{l+1}$ and $v.$ Since $x_l\prec y_{l+1},$ we
conclude that $\{x_l,u\}\prec \{y_{l+1},v\},$ yielding  $C_l\prec
C_{l+1}.$ Now, consider the cell $C_{l+2}.$ The element $y_{l+2}$
must be located to the right of $x_{l+1},$ therefore to the right of
$C_l.$ Since $C_{l+2}$ and $C_{l}$ belong to different blocks, we
can show that $C_l\prec C_{l+2}$ by using exactly the same reasoning
as for the cells $C_l$ and $C_{l+1}.$ Continuing this way, we obtain
the required relationship $C_l\prec C_{l+1}\cup\cdots\cup C_{l+r}.$
This establishes the assertion. Suppose that
$[1,i_1],[i_1+1,i_2],\ldots,[i_r+1,k]$ are the indices of cells
defining the beginning and the end of each group. From our assertion
we infer that $C_k\prec C_{i_1}\prec C_{i_2}\prec\ldots\prec
C_{i_r}\prec C_k,$ contrary that $\prec$ is a total order.
\end{proof}

\begin{lemma}\label{G3_path} If $C\rightarrowtail C'$ is a
(G3)-arc and $C$ belongs to a principal component, then  $C$ and
$C'$ belong to the same cluster. In particular, ${\mathcal G}_{ij}$
does not contain (G3)-cycles or no $\epsilon$-compatible order
exist. Moreover, ${\mathcal G}_{ij}$ does not contain (G2)-cycles.
\end{lemma}

\begin{proof} Let $xy$ be a strongly separated pair with $x\in
C.$ Since  $C\rightarrowtail C'$ is a (G3)-arc, there exist $y'\in
C$ and $x'\in C'$ such that $y'\rightarrowtail x'$ is an (A2)-arc.
Then there exists  $z'$ such that $x'z'$ is strongly separated. If
$xz$ and $x'y'$ belong to different principal components, then there
exists a (G2)-arc from $C'$ to $C$ or from $C$ to $C'$. In the first
case, $C$ and $C'$ obey {\bf (S2)}, thus we cannot have a (G3)-arc
from $C$ to $C'.$ Analogously, in the second case, we deduce that we
have at the same time a (G3)-arc and a (G2)-arc from $C$ to $C'.$
This is impossible, so $C$ and $C'$ belong to a common principal
component. Now, if
${\mathcal G}_{ij}$ contains a (G3)-cycle, then the first assertion
implies that all its cells belong to the same cluster, and Lemma
\ref{cycle} yields that no $\epsilon$-compatible order exist.
Finally, let ${\mathcal C}=(C_1,C_2,\ldots,C_k,C_1)$ be
a (G2)-cycle.  In each  $C_i,$ pick  $x_i,y_i$ so that
$d_{x_i} \ll_4 d_{y_{i+1 (mod k)}}.$ Since there is no {\bf (G2)} or {\bf (G3)}
arc from $C_{i+1 (mod k)}$ to $C_i,$ we get
$d_{y_{i}} \lesssim_4 d_{x_{i+1 (mod k)}},$
yielding $d_{x_i} \ll_4 d_{y_{i+1 (mod k)}} \lesssim_4 d_{x_{i+2
(mod k)}}.$ Thus $d_{x_i} < d_{x_{i+2 (mod k)}}$ for $i=1,\ldots k.$
Then $d_{x_1}<d_{x_3}<\cdots <d_{x_{k-1}}<d_{x_1}$ for even $k$  and
$d_{x_1}<d_{x_3}<\cdots <d_{x_k}<d_{x_2}<d_{x_4}<\cdots
<d_{x_{k-1}}<d_{x_1}$ for odd $k,$ a contradiction. \end{proof}

To complete the bipartition of  cells into two acyclic subgraphs of
${\mathcal G}_{ij}$, it remains to deal with induced mixed cycles.
The following results precise their structure.

\begin{lemma}\label{G2} Any induced mixed cycle $\mathcal C$ of
${\mathcal G}_{ij}$ contains one or two (G2)-arcs, and if $\mathcal
C$ contains  two such arcs, then they are consecutive.
\end{lemma}

\begin{lemma} \label{G3G2} Let $C'\rightarrowtail C$ be a (G3)-arc,
$C\rightarrowtail C''$ be
a (G2)-arc, and suppose that there is no (G2)-arc from $C'$ to
$C''.$ If $C,C'$ do not belong to distinct twin clusters and
$C,C''$ do not belong to the same cluster, then $C$ and $C'$ must
be separated.
\end{lemma}

Thus a mixed cycle $\mathcal C$ contains either one (G2)-arc
($\mathcal C$ is a 1-{\it
cycle}) or two consecutive (G2)-arcs ($\mathcal C$ is a
2-{\it cycle}), all other arcs of $\mathcal C$ being (G3)-arcs.
By Lemma \ref{G3_path}, the heads of all (G3)-arcs of $\mathcal C$
are (G3)-cells of the same cluster $\mathcal K$. Then we
say that the cycle $\mathcal C$ {\it intersects} the cluster
${\mathcal K}.$ For a (G2)-arc  $C_0\rightarrowtail C$ and a cluster
$\mathcal K$, we show how to detect if there exists a 1-
or 2-cycle $\mathcal C$ passing via $C_0\rightarrowtail C$ and
intersecting $\mathcal K$. We  consider the case of 1-cycles. Then
$C_0$ must be a (G3)-cell of ${\mathcal K}.$ Note that
an induced 1-cycle cannot contain cells $C'$
such that $C_0\rightarrowtail C'$ is a {\bf (G2)} or
{\bf (G3)}-arc. Hence, we can remove all such cells of $\mathcal K$.
Analogously, we remove all cells $C'$ so that
$C'\rightarrowtail C$ is an arc. In the subgraph induced by the
remaining cells of ${\mathcal K}$ we search for a shortest directed
path ${\mathcal Q}=C\rightarrowtail C_1\rightarrowtail \cdots
\rightarrowtail C_k\rightarrowtail C_0$ so that the first arc
$C\rightarrowtail C_1$ and the last arc $C_k\rightarrowtail C_0$ of
this path are (G3)-arcs. This can be done in polynomial time by
testing all possible choices for $C_1$ and $C_k$ and applying for
each pair a shortest path finding algorithm in an acyclic graph. If
such a path $\mathcal Q$ does not exist, then no required induced
cycle $\mathcal C$ exist. Otherwise, the path
$\mathcal Q$ together with the arc $C_0\rightarrowtail C$ define an
induced cycle $\mathcal C$ having exactly one (G2)-arc. Indeed, if
$C_i\rightarrowtail C_j$ is a {\bf (G2)} or {\bf (G3)}-arc
and $|i-j|>2$, since the subgraph induced by $\mathcal K$ is
acyclic, we must have $i<j.$ This contradicts the minimality of the
path ${\mathcal Q}.$ So, the resulting cycle is indeed induced. It remains to
note that $\mathcal C$ does not contain other (G2)-arcs, because
by Lemma \ref{G2} in an induced cycle the (G2)-arcs are consecutive.
Analogously, we can decide  if there exists a 2-cycle passing via
$C_0\rightarrowtail C$ and intersecting $\mathcal K$, and having a
second (G2)-arc of the form $C\rightarrowtail C'_0$ or
$C'_0\rightarrowtail C_0.$ Therefore, we have the following result:

\begin{lemma} \label{induced-mixed} For a (G2)-arc
$C_0\rightarrowtail C$ and a  cluster ${\mathcal K}$, one can decide
in polynomial time if there exists an induced 1- or 2-cycle
$\mathcal C$ passing via $C'\rightarrowtail C$ and intersecting
${\mathcal K}.$
\end{lemma}

For a cell $C$, let $\Omega_1(C)$ be the set of (G2)-arcs
$C_0\rightarrowtail C$ belonging to a 1-cycle intersecting a
cluster $\mathcal K$ not containing $C$. Let $\Omega_2(C)$ be the
set of (G2)-arcs $C_0\rightarrowtail C$ belonging to  a 2-cycle
$\mathcal C$ intersecting a cluster $\mathcal K$ not containing $C$
and passing via $C_0\rightarrowtail C$ so that the arc of $\mathcal
C$ entering $C_0$ is a (G3)-arc. In both cases $C_0$
belongs to $\mathcal K$: $C_0$ is a head of a (G3)-arc of $\mathcal
C$, and all such heads belong to $\mathcal K$. Finally, let
$\Omega_3(C)$ be the set of (G2)-arcs $C\rightarrowtail C_0$ belonging
to a 2-cycle $\mathcal C$ intersecting a cluster $\mathcal
K$, so that $C$ belongs to $\mathcal K$ and the arc of $\mathcal C$
entering $C$ has type {\bf (G2)}. Fig. 2 illustrates this classification.
For each cell $C$ of ${\mathcal
G}_{ij}$ we introduce a binary variable $x_C$ satisfying the
following constraints: {\bf (F1)} $x_{C'}=x_{C''},$ if  $C',C''$ belongs to the
same cluster; {\bf (F2)} $x_{C'}\ne x_{C''},$ if $C',C''$ belong to twin
clusters; {\bf (F3)} $x_C\ne x_{C_0},$ if the arc
$C_0\rightarrowtail C$ belongs to $\Omega_1(C)\cup \Omega_2(C);$
{\bf (F4)} $x_C\ne x_{C_0},$ if the arc $C\rightarrowtail
C_0$ belongs to $\Omega_3(C)$. Define a 2-SAT formula $\Phi_{ij}$
by replacing every constraint
$a=b$ by two clauses $(a\vee \bar{b})$ and $(\bar{a}\vee b)$ and
every constraint  $a\ne b$ by two clauses $(a\vee b)$ and
$(\bar{a}\vee \bar{b})$.

\begin{figure}
\begin{center}
\scalebox{0.40}{\input{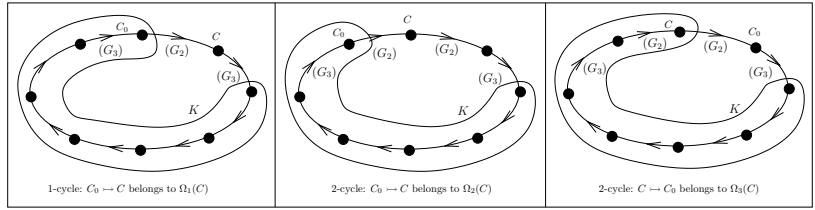}}
\end{center}
\caption{To the classification of the arcs incident to a cell $C$}
\end{figure}

\begin{proposition} \label{acyclic} If the 2-SAT formula $\Phi_{ij}$
admits a satisfying assignment $A$, then the sets $X^-_{ij}=\{ C:
A(x_C)=0\}$ and $X^+_{ij}=\{ C: A(x_C)=1\}$ define a partition of
${\mathcal G}_{ij}$ into two acyclic subgraphs. Conversely, given an
$\epsilon$-compatible order on $X,$ the assignment $A$ defined by
setting $A(x_C)=0$  if $C$ is located in
 $H_i,$  $A(x_C)=1$ if $C$ is located in  $H_{j-1},$ and
$A(x_{C'})=A(x_{C''})$ if $C'$ and $C''$ are located in a common
inner hole,  is a true assignment for $\Phi_{ij}.$ In particular, if
$\Phi_{ij}$ is not satisfiable, then no $\epsilon$-compatible order
exist.
\end{proposition}

\begin{proof}
Let $A$ be a true assignment  of $\Phi_{ij}$ and the
partition $X^-_{ij},X^+_{ij}$ of $X_{ij}$ be defined as above.
Denote by ${\mathcal G}^-_{ij}$ and ${\mathcal G}^+_{ij}$ the
subgraphs induced by $X^-_{ij}$ and $X^+_{ij}.$ {\bf (F1)}
forces every cluster to be included in one set. {\bf (F2)}
implies that the twin clusters are separated. Hence ${\mathcal
G}^-_{ij}$ and ${\mathcal G}^+_{ij}$ do not contain (G1)-cycles: if
$C$ and $C'$ are the two cells of a (G1)-cycle, then
$(x_C\vee x_{C'})\wedge ({\bar x_C}\vee {\bar x_{C'}})$ yields
$A(x_C)\ne A(x_{C'}).$ By Lemma \ref{G3_path}, ${\mathcal G}_{ij}$
does not contain (G2)-cycles. Since the cells of a (G3)-cycle are
contained in the same cluster and each cluster
induces an acyclic subgraph, ${\mathcal G}^-_{ij}$ and ${\mathcal
G}^+_{ij}$ do not contain (G3)-cycles as well. Now, let ${\mathcal
G}^+_{ij}$ contain a mixed cycle. Then it also contains an induced
mixed cycle ${\mathcal C}.$ From Lemma \ref{G2} we infer that
$\mathcal C$  has either one (G2)-arc $C_0\rightarrowtail C$ or
exactly two consecutive (G2)-arcs $C_0\rightarrowtail
C\rightarrowtail C''$. In the first case, we conclude that
$C_0\rightarrowtail C$ belongs to $\Omega_1(C),$ thus
 {\bf (F3)} yields $x_C\ne x_{C_0}$, contrary to the fact that
$A(x_C)=A(x_{C'})=1.$ Analogously, in the second case, we deduce
that either $x_C\ne x_{C_0}$ and the arc $C_0\rightarrowtail C$
belongs to $\Omega_2(C)$ or $x_C=x_{C_0}$ and the arc
$C\rightarrowtail C''$ belongs to $\Omega_3(C),$ whence $x_C\ne
x_{C''}.$ Then we obtain a contradiction with the assumption that
$A(x_{C_0})=A(x_C)=A(x_{C''})=1.$  This shows that the subgraphs
${\mathcal G}^-_{ij}$ and ${\mathcal G}^+_{ij}$ obtained from the
true assignment $A$ of $\Phi_{ij}$ are acyclic.

Conversely, let $A$ be an assignment obtained from an
$\epsilon$-compatible order as defined in the proposition. We assert
that $A$ is a true assignment for $\Phi_{ij},$ i.e., it satisfies
the constraints {\bf (F1)-(F4)}. This is obvious for constraints
{\bf (F1)} and {\bf (F2)}, because if two cells $C',C''$ belong to
the same cluster, then they will be located in the same hole and we
must have $A(x_{C'})=A(x_{C''}).$ If $C'$ and $C''$ belong to
distinct twin clusters, then they must be separated, therefore the
unique $\epsilon$-admissible location of $C'$ and $C''$ will be in
different bounding holes, thus  $A(x_{C'})\ne A(x_{C''}).$ Now, pick
an arc $C_0\rightarrowtail C$ which belongs to $\Omega_1(C)\cup
\Omega_2(C).$ If $C_0\rightarrowtail C$ belongs to $\Omega_1(C),$
then there exists a 1-cycle $\mathcal C$ passing via
$C_0\rightarrowtail C$ and intersecting a cluster $\mathcal K$.
Since all cells of $\mathcal C$, except $C$, are heads of (G3)-arcs,
they all belong to $\mathcal K$, i.e., they  have the same value in
the assignment. By Lemma \ref{cycle}, $C$ must be separated from
$C_0$ (namely $C$ and $C'$ must be located in different bounding
holes), showing that $A(x_C)\ne A(x_{C_0})$. If  $C_0\rightarrowtail
C$ belongs to $\Omega_2(C),$ then let $\mathcal C$ be a 2-cycle
passing via $C_0\rightarrowtail C$ and intersecting the cluster
$\mathcal K$ not containing $C.$ Additionally, we know that the arc
$C'\rightarrowtail C_0$ of $\mathcal C$ entering $C_0$ is a
(G3)-arc, thus $C_0$ belongs to $\mathcal K.$  Since $C'$ cannot
belong to the twin cluster of ${\mathcal K}$ (this will contradicts
that $C'\rightarrowtail C_0$ is a (G3)-arc) and since $C$ does not
belong to $\mathcal K,$ from Lemma \ref{G3G2} we infer that  $C_0$
and $C$ are separated, thus $A(x_C)\ne A(x_{C_0})$. Finally, let
$C\rightarrowtail C_0$ belong to $\Omega_3(C).$ Then there exists a
2-cycle $\mathcal C$ passing via $C\rightarrowtail C_0$ and
intersecting the cluster $\mathcal K$, such that $C$ belongs to
$\mathcal K$ and the arc of $\mathcal C$ entering $C$ has type {\bf
(G2)}. Since all cells of $\mathcal C$ except $C$ and $C_0$ are
heads of (G3)-arcs, they all belong to $\mathcal K$. Since $C$ also
belongs to this cluster, by Lemma \ref{cycle}, $C_0$ must be
separated from the remaining cells of $\mathcal C$, yielding $x_C\ne
x_{C_0}.$ Hence $A$ satisfies the constraints {\bf (F1)-(F4)}. This
shows, in particular, that if $\Phi_{ij}$ is not satisfiable, then
no $\epsilon$-compatible order exist.
\end{proof}

\medskip\noindent{\sc\bf  4.3. Sorting the cells of $X^-_{ij}$ and
$X^+_{ij}.$} Let ${\mathcal G}^-_{ij}$ and ${\mathcal G}^+_{ij}$ be
the subgraphs of  ${\mathcal G}_{ij}$ induced by the sets
$X^-_{ij}$ and $X^+_{ij}$ obtained from the true assignment of the
2-SAT formula $\Phi_{ij}.$ We will locate all cells of $X^-_{ij}$ in
the hole $H_i$ and all cells of $X^+_{ij}$ in the hole $H_{j-1}$ of
$H_{ij}.$ The elements from two cells $C',C''$ located in the same
hole will not be mixed, i.e., $C'$ will be placed to
the right of  $C'',$ or vice versa.  To specify the
total order among cells, we use that ${\mathcal
G}^-_{ij}$ and ${\mathcal G}^+_{ij}$ are  acyclic, therefore each of
them admit a topological order.  We compute a
topological order $C_{j_1} \prec C_{j_2} \prec \ldots \prec C_{j_p}$
on the cells of $X^+_{ij}$ and  a dual topological order $C_{i_q}
\prec C_{i_{q-1}} \prec \ldots \prec C_{i_1}$ on the cells of
$X^-_{ij}$. We locate the cells of $X^+_{ij}$ in $H_{j-1}$ and the
cells of $X^-_{ij}$ in $H_{i}$ according to these orders. The
following  two results relay the topological orders on the cells
with the order on the distances between
elements from such cells.

\begin{lemma} \label{ordre-C_i} Let $C',C''$ be two cells
of $X^+_{ij}.$ If $C' \prec C''$ in the topological order, then for
any   $y\in C'$, $z\in C''$ and $x\in X^-_{ij}$, we have $d_y
\lesssim_4 d_z$ and $d(x,y) \lesssim_{16} d(x,z).$
\end{lemma}

\begin{proof} Since $C',C''$ belong to $X^+_{ij},$ they
are not connected by (G1)-arcs. Since $C' \prec C''$ in the
topological order, there is no arc from $C''$ to $C'$. As
$C''\rightarrowtail C'$ is not a (G2)-arc, we must
have $d_{z}\gtrsim_4 d_y.$ As $C''\rightarrowtail C'$
is not a (G3)-arc, we obtain  $d(x,y) \lesssim_{16} d(x,z).$
\end{proof}

\begin{lemma} \label{ordre-C_ii} Let  $C,C',C''$ be three distinct
cells of the graph ${\mathcal G}_{ij}.$   If the algorithm returns
the total order $\prec$ and $C\prec C'\prec C'',$ then for any  $x
\in C, y \in C', z \in C''$ or $x,y,z \in C\cup C'$ and $x\prec
y\prec z,$ we have $d(x,z) \gtrsim_{16} \max \{d(x,y),d(y,z)\}.$
\end{lemma}

\begin{figure}
\begin{center}
\scalebox{0.7}{\input{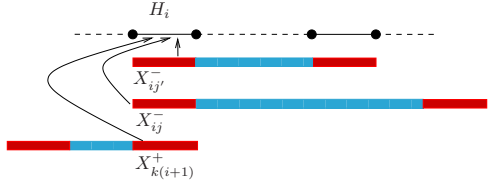}}
\end{center}
\vspace*{-0.4cm} \caption{Relative location of the cells of
$X_{k(i+1)}^+,X_{ij'}^-,$ and $X_{ij}^-$ $(k<i,j'<j)$ in
$H_i$} \vspace*{-0.3cm}
\end{figure}

After fixing the relative position of each cell $C$ of $X_{ij},$ we
make a recursive call to  $C.$  For this, we update the canonical
order $\preccurlyeq$ in the following way: if $C$ is located in
$X^+_{ij},$ we set $x\preccurlyeq^+ y$ if $x\rightarrowtail y,$
otherwise, if $C$ is located in $X^-_{ij},$ we set
$x\preccurlyeq^- y$ if $y\rightarrowtail x.$ Since
$\preccurlyeq^+$ and $\preccurlyeq^-$ are dual, if we
apply to them the ``closing" rules,  we will obtain two dual
partial orders, denoted also by $\preccurlyeq^+$ and
$\preccurlyeq^-.$ The restriction on $C$ of every
$\epsilon$-compatible order $\prec$ on $X$ is an extension of
$\preccurlyeq^+$ or $\preccurlyeq^-:$ since all elements of $C$
will be placed in the same hole, either $a_{i+1} \prec C$  or  $C \prec a_j.$ If $a_{i+1}
\prec C$,  then $x\prec y$ for all $x,y\in
C$ such that $x\rightarrowtail y.$ Hence
$\prec$ is a linear extension of $\preccurlyeq^+$. Therefore, if the
recursive call to a cell $C$ returns the answer ``not", then  no
$\epsilon$-compatible total order on $X$ exist. Else,  it
returns a total order on $C$, which is $16\epsilon$-compatible by
induction hypothesis. Then, the total order between the
cells of ${\mathcal G}_{ij}$  and the total orders on cells are
concatenated to give a single total order $\prec$ on $X_{ij}.$

\medskip\noindent{\sc\bf  4.4. Defining the total order on $X_i$.}
Recall that $X_i$ is the set of all elements of $X^{\circ}$
located in the hole $H_i.$ According to our algorithm,  $X_i$ is
the disjoint union of all sets $X^-_{ij}$ $(j>i+1)$ and
$X^+_{k(i+1)}$ $(k<i)$.  We just defined a total order between the
cells of each of the sets $X^-_{ij},X^+_{k(i+1)},$ and applying
recursion we defined a total order on the elements of each cell.
To obtain a total order on the whole set $X_i$ it remains to
define a total order between the sets $X^-_{ij}$ $(j>i+1)$ and
$X^+_{k(i+1)}$ $(k<i).$  For this, we locate each $X^+_{k(i+1)}$
$(k<i)$ to the left of each $X^-_{ij}$ $(j>i).$ Given two sets
$X^+_{k(i+1)}, X^+_{k'(i+1)}$ $(k,k'<i)$, we locate $X^+_{k(i+1)}$
to the left of $X^+_{k'(i+1)}$ if and only if $k<k',$ i.e., iff
$H_{k(i+1)}\Subset H_{k'(i+1)}.$ Analogously, given $X^-_{ij},
X^-_{ij'}$ $(j,j'>i+1)$, we locate $X^-_{ij'}$ to the right of
$X^-_{ij}$ if and only if $j'<j,$ i.e., iff $H_{ij'}\Subset
H_{ij}.$ This location is justified by the Proposition
\ref{pairwise_admissible} and is illustrated in Fig. 3.

\vspace{-0.4cm}
\section{The algorithm and its performance guarantee}
We have collected all necessary tools to describe the algorithm. It consists  of three procedures
$l_{\infty}$-{\sf Fitting\_by\_Robinson}, {\sf Refine}, and {\sf
Partition\_and\_Sort}. The main procedure $l_{\infty}$-{\sf
Fitting\_by\_Robinson} constructs the sorted list $\Delta$ of
feasible values for the optimal error $\epsilon^*$. Its entries are
considered in a binary search fashion and the algorithm returns the
smallest value $\epsilon\in \Delta$ occurring in this search for
which  the answer ``not'' is not returned (i.e., the least
$\epsilon$ for which a $16\epsilon$-compatible total order on $X$
exists). To decide, if, for a given $\epsilon$, such an order
exists, the procedure {\sf Refine}$(X,\preccurlyeq, \epsilon)$
constructs (and/or updates) the canonical partial order
$\preccurlyeq$ and computes a maximal chain $P$ of
$(X,\preccurlyeq).$ For each element $x\in X^{\circ}:=X\setminus P,$
{\sf Refine} computes the set $AH(x)$ of all $x$-holes which
participate in $(x,y,1)$-admissible locations for all $y\in
X^{\circ}$ and defines the segment $H(x)$. For each pair $i<j-1,$
{\sf Refine} constructs the set $X_{ij}$ and makes a call of the
procedure {\sf Partition\_and\_Sort}$(X_{ij}),$ which returns the
bipartition $\{X^-_{ij},X^+_{ij}\}$ of $X_{ij}$ and a total order on
the cells of $X^-_{ij}$ and $X^+_{ij}.$ Then {\sf Refine}
concatenates in a single total order on cells the total orders on
cells coming from different  sets assigned to the same hole. After
this, {\sf Refine} is recursively applied to each cell occurring in
some graph ${\mathcal G}_{ij}.$ The returned total orders on cells
are concatenated into a single total order $\prec$ on $X$ according
to the total orders between cells and between holes; then $\prec$ is
returned by the algorithm $l_{\infty}$-{\sf Fitting\_by\_Robinson}. The procedure
{\sf Partition\_and\_Sort} constructs the graphs
${\mathcal L}_{ij}$ and ${\mathcal L}_{ij}^{\rightarrow}.$ Using
these graphs, $X_{ij}$ is partitioned into blocks and cells, then
graph ${\mathcal S}_{ij}$ and its clusters are
constructed. Using the cells, the directed graph ${\mathcal
 G}_{ij}$ is constructed. If ${\mathcal S}_{ij}$ is not bipartite or ${\mathcal
 G}_{ij}$ contains (G3)-cycles, then {\sf
 Partition\_and\_Sort} returns the answer ``not''. Otherwise, for
 each cell $C$ and each cluster $\mathcal K$, it tests if there exists a 1-cycle
 and/or a 2-cycle passing via $C$ and intersecting $\mathcal K$. Consequently,
 for each cell $C$, the lists $\Omega_1(C),\Omega_2(C),$ and $\Omega_3(C)$ of
 (G2)-arcs are computed. These lists are used to construct the 2-SAT
 formula $\Phi_{ij},$ which is solved by the algorithm of
 \cite{AsPlTa}. If $\Phi_{ij}$ admits a true assignment $A,$ then
 $X^-_{ij}=\{ C: A(x_C)=0\}$  and $X^+_{ij}=\{ C: A(x_C)=1\}$ define
 a bipartition of $X_{ij}$ into two acyclic subgraphs ${\mathcal G}^-_{ij},
 {\mathcal G}^+_{ij}$ of
 ${\mathcal G}_{ij}.$ Then {\sf
 Partition\_and\_Sort}  locates  the cells from
 $X^+_{ij}$ in the  hole $H_{j-1}$  according to
 the topological order of the acyclic
 graph  ${\mathcal G}^+_{ij}$ and it
 locates the cells from $X^-_{ij}$ in the hole $H_i$ according to the dual
 topological order of ${\mathcal
 G}^-_{ij}.$ Note that if at some stage
 {\sf Refine} or {\sf Partition\_and\_Sort}
 returns the answer ``not", then there does not exists any
 $\epsilon$-compatible total order on $X$ and the current value
 of $\epsilon$ is too
 small. The total complexity of the algorithm  is $O(n^6\log n).$ We formulate now the main result of our paper:

\begin{theorem}
\label{theorem} For $\epsilon\in \Delta$, if the algorithm returns
the answer ``not", then the dissimilarity $d$ is not
$\epsilon$-Robinson, else, it returns a $16\epsilon$-compatible
total order $\prec$ on $X.$ In
particular, the algorithm is a factor 16 approximation algorithm for
$l_{\infty}$-{\sc FITTING-BY-ROBINSON}.
\end{theorem}

\begin{proof} First, note that no $\epsilon$-compatible order exist in all
cases when the algorithm returns the answer ``not". Indeed, Lemma
\ref{G3_path}, Propositions 3.2 and 4.10 cover all such cases except the case when
this answer is returned by a recursive call. In this case, the
induction assumption implies that no $\epsilon$-compatible total
order on $C$ extending $\preccurlyeq^+$ (and therefore its dual
$\preccurlyeq^-$) exist. Then  we infer that no
$\epsilon$-compatible order on $X$ exist as well.

Now,
let the algorithm return a total order $\prec.$ Suppose by
induction assumption that $\prec$ is $16\epsilon$-compatible on each
cell to which  a recursive call is applied. On the chain $P,$ the
total order  $\prec$ coincides with $\preccurlyeq$, therefore
$\prec$ is $\epsilon$-compatible on $P.$ Moreover, $\prec$ is
$\epsilon$-compatible on $P\cup \{ x\}$ for any $x\in X^{\circ},$
because every element $x$ is located in a bounding hole of $H(x)$
which is $x$-admissible. Finally notice that $\prec$ is
$12\epsilon$-compatible on $P\cup\{ x,y\}$ for any $x,y\in
X^{\circ}$  because by Proposition \ref{pairwise_admissible} the
bounding hole of $H(x)$ and the bounding hole of $H(y)$ into which
$x$ and $y$ are located define a $(x,y,12)$-admissible pair. To
prove that $\prec$ is $16\epsilon$-compatible on the whole set
$X,$ it suffices to show that $d(x,z)\gtrsim_{16}
\mbox{max}\{d(x,y),d(y,z)\}$ for any three elements $x,y,z\in X$
such that $x\prec y\prec z.$ From previous discussion, we can
suppose that $x,y,z\in X^{\circ}$.  For this, we distinguish the
Cases {\bf (H1)-(H5)} in function of the mutual location of segments
$H(x)$ and $H(z)$ and in each case we show the required inequality.
The respective case analysis is given in \cite{ChSe}.

\end{proof}

\vspace*{-0.4cm}

\begin{footnotesize}

\end{footnotesize}

\vspace*{-1.5cm}
\end{document}